# Amplification-free GW-level, 150 W, 14 MHz, and 8 fs, thin-disk laser oscillator

SEMYON GONCHAROV,*[1] KILIAN FRITSCH[2], OLEG PRONIN[1]

[1]Helmut Schmidt University / Universität der Bundeswehr Hamburg, Holstenhofweg 85, 22043 Hamburg, Germany
[2]n2-Photonics, Hans-Henny-Jahnn-Weg 53, 22085 Hamburg, Germany

*Corresponding author: semyon.goncharov@hsu-hh.de

**We report an amplification-free thin-disk laser oscillatory system delivering 0.9 GW peak power. The 120 fs pulses at 14 MHz containing 12.8 µJ delivered by thin-disk oscillator were compressed by factor 15 down to 8.0 fs with 148 W average output power and 82 % overall efficiency. Additionally, we showed that even a sub-two-cycle operation with 6.2 fs can be reached with this technology. The system will be a crucial part of the XUV frequency comb being developed and a unique high-repetition rate driver for attosecond pulse generation.**

Spectral broadening and compression of ultrashort pulses are widely used to generate few-cycle electric fields. These few-cycle oscillations of light have proven to be advantageous for a wide range of applications like isolated attosecond pulse generation [1], ultrafast pump-probe spectroscopy [2], nanoscopy [3], attosecond science [4], and X-ray or XUV sources [5,6]. Since its first demonstration [7], Herriott-type multipass cells have become a rapidly developing technique for high broadening and compression factors. A multipass cell represents a quasi-waveguide system where a laser beam can propagate distances much longer than the cell's geometrical size. Thus, the extended interaction distance of a pulse with a nonlinear medium [8], geometrical simplicity, low sensitivity to beam pointing, and advantage of average power scalability [9–11] favor its adoption and applicability. However, despite the rapid advancement of this technique, only a few demonstrations approached sub-10 fs pulse durations [10,12–16]. This is due to the challenges of supporting the necessary bandwidth and managing the corresponding mirrors' dispersion, which, in turn, compromise the quality of pulse compression. In contrast to waveguides exploiting the total internal reflection, and thus, in principle, not limited to the spectral bandwidth, the multipass cells require mirrors that should have metal or dielectric coatings. Metal coatings have the big advantage of being extremely broadband and dispersion-free; however, showing significant absorption losses on the order of a few percent per reflection. Dielectric mirrors can approach one octave [17,18] bandwidth, have losses on the order of ~0.1 %, and additionally show the oscillations of group delay dispersion (GDD). Importantly, even though fibers can support extreme bandwidths, efficient compression down to a few cycles and below requires excellent dispersion control. This, in turn, requires dispersive mirrors, which can compensate for higher-order dispersion terms. Such mirrors, for example, were implemented in the light-field synthesizers [19]. Applying a similar concept with different spectral channels broadened and compressed in a multipass cell is possible, but it can be complex and practically challenging to implement.

Among the few-cycle laser systems, high average power multi-megahertz repetition rate sources are highly desirable for ultrafast XUV spectroscopy applications [20,21]. However, to date, these kinds of lasers are rather underrepresented. Most existing amplification-based systems can deliver few-cycle pulses and operate in the 1 kHz-1 MHz range [5,22,23]. Thin-disk oscillators offer a unique combination of high average power and high peak power [24], simultaneously serving as a most straightforward femtosecond system, mode-locked oscillator. However, the pulse duration and peak powers reached are insufficient for many applications. Here, we show that the peak power of the oscillator can be boosted to GW-level and simultaneously provide a high average power of 150 W in combination with a high repetition rate of 14 MHz and two-cycle pulse duration <8 fs. On the other hand, Herriott cells proved to be robust with regard to misalignment-triggered damage while maintaining an excellent output beam-[25] and pulse quality [14], high throughput (>90 %), and compact portable form factor. Therefore, multipass cells based on dielectrically coated mirrors can be a great candidate to enable an ultrafast laser source that combines multi-megahertz repetition rate, high average power, sub-10 fs pulses, and long-term power- and pointing stability, which in turn would noticeably alleviate challenges for XUV frequency comb spectroscopy.

Previously, we demonstrated spectral broadening and pulse compression in cascaded multipass cells based on dispersive dielectric mirrors (spanning from 850-1350 nm) and Argon as a nonlinear medium for a relatively low-average-power laser [14]. We experimentally showed that with proper intra-cell dispersion management, white light can be generated by operating in an anomalous dispersion regime. In this work, we applied the same concept to our previously developed high-average- and peak-power oscillator [26]. The experimental setup is shown in Fig. 1. The driving laser represented a home-built Kerr-lens mode-locked oscillator delivering 120 fs long pulses at 14 MHz containing 12.8 µJ energy, resulting in a 180 W average power. The oscillator output was mode-matched to the stage 1 eigenmode. The first Herriott cell was assembled in a monolithic aluminum housing and consisted of two highly reflective mirrors (150 mm radius of curvature each) separated by 136 mm. A 3-mm thick AR-coated fused silica plate was placed 5 mm away from the cell mirror, and a 1-mm thick AR-coated fused silica plate was placed in the center of the cell. Both plates served as nonlinear media. After 34 passes through the cell, a total B-integral of ~27 rad was accumulated, corresponding to ~0.8 rad per pass. This multipass cell operated in an ambient air environment. The output pulses were compressed to 38 fs [see Fig. 2] with 8 bounces off dispersive mirrors, providing each -400 $fs^2$ of GDD per bounce. The commercial SPIDER device (APE GmbH) verified the compression. Considering 90 % of the energy in the main peak, the output pulses carried 250 MW peak power with 169 W average power. The transmission of the first stage remained at 94 % and was mainly attributed to losses in the fused silica plates. The inset in Fig. 2(b) shows the output beam. Additionally, we carried out a one-dimensional propagation simulation of the seed pulses through stage 1. Considering self-steepening and self-phase modulation in this cell, it was possible to predict the resulting broadening reliably [see Fig 2(a)]. It should be noted that the losses introduced by the compressor mirrors were negligible.

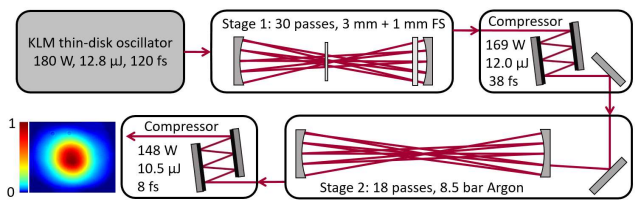

**Fig. 1.** Schematic setup of the nonlinear broadening and pulse compression.

Additionally, we characterized spatio-spectal homogeneity by collecting optical spectra in different positions across the stage 1 output beam. The beam was coupled into a spectrometer via a 200 um-core-size multimode fiber. The overlap parameter V [25] was used to quantify the measurement [see Fig. 3]. The weighted mean values are $V_x$ = 99.5 % and $V_y$ = 99.5 %, respectively, indicating an excellent spectral overlap across the area of the beam. The beam quality parameter of the stage 1 output was measured following the ISO 11146 procedure and was found to be 1.18.

The compressed pulses were coupled via a mode-matching mirror to the second stage in the following step. The second Herriott cell included a complementary pair of dispersive mirrors with a 200 mm radius of curvature separated by 388 mm. The assembly was placed in a monolithic aluminum housing filled with Argon. The cell configuration included 18 passes through the gas volume. The cell mirrors were designed to provide approximately -90 $fs^2$ of GDD after two bounces around 1030 nm. Thus, the cell could operate in positive and negative dispersive regimes by fine-tuning the gas pressure.

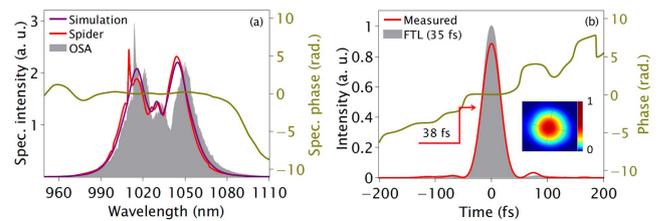

**Fig. 2**. Temporal characterization of the stage 1 output. (a) The spectrum phase and intensity were retrieved with a commercial SPIDER device. The spectrum was additionally measured with an optical spectrum analyzer (OSA). The purple curve represents the simulated output spectrum. (b) Temporal phase and intensity compared to Fourier transform limit of the spectrum (FTL). The main peak included 90 % of the pulse energy. The output beam profile is shown in the inset.

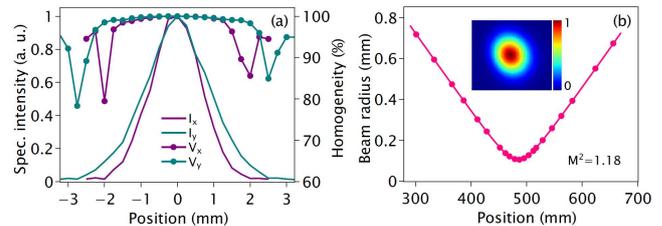

**Fig. 3.** Characterization of the stage 1 output. (a) Spatio-spectral homogeneity V and relative intensity I for tangential and sagittal planes. (b) Beam quality measurement $M^2$ according to ISO 11146 assuming a 1032 nm central wavelength. The dots and the lines represent experimental data and fit, respectively. The beam in the focal plane is depicted in the inset.

Considering the group velocity dispersion of Argon as 0.015 $fs^2$/mm [27], a pressure of ~7.8 bar was estimated as a transitional value between negative and positive dispersive regimes. Firstly, the Argon pressure was adjusted to ~8 bar, ensuring a slightly positive dispersive regime. That allowed us to fully utilize the entire bandwidth of the broadband mirrors (850 – 1300 nm) without sacrificing the overall transmission of the system. The second stage output spectrum is shown in Fig. 4.

The pulse duration and the phase information were retrieved with the commercial SPIDER device [see Fig. 4]. The main peak contained 75 % energy or ~0.87 GW peak power based on the retrieval. These values correspond to factor 15 of temporal compression and factor 9 of peak power increase. The second stage throughput in the positive dispersion regime was 88 %, essentially defined by the cell mirror coatings, while the overall transmission after both stages and the compressors remained 82 %. The output spectrum was measured by an optical spectrum analyzer from Ando AQ6317B. The Fourier transform limit of 7.8 fs was calculated from the measured spectrum. The output pulses were compressed with 4 bounces off dispersive mirrors (4 x -45 fs$^2$) and a 6 mm thick AR-coated fused silica window down to 8.0 fs. The quality of pulse compression can be potentially improved by introducing higher order dispersion terms in the compressor mirrors.

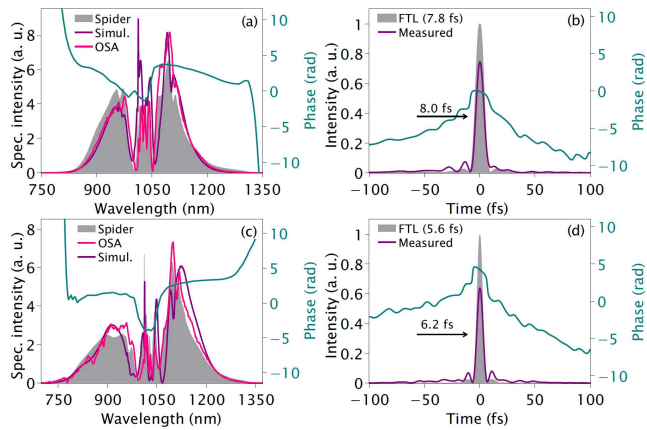

**Fig. 4.** Temporal characterization of the stage 2 output. (a, c) Spectral phase and intensity were retrieved with a commercial SPIDER device. The spectrum was additionally measured with an optical spectrum analyzer (OSA). The purple curves represent the simulated output spectrum. (b, d) Temporal phase and intensity are compared to the spectra's Fourier transform limit (7.8 fs and 5.6 fs, respectively). The main peak included 75 % of the pulse energy in (b) and 65 % in (d).

The nonlinear phase shift/B-integral in the second stage was estimated to be 0.8 rad per pass. No significant deterioration of the output beam was observed [see Fig. 1 and Fig. 5(c)]. However, the sensitivity of our Si-based CMOS sensor spanned the range of 320 – 1100 nm (Cinogy), which allowed us to verify the beam quality after the second stage only within this spectral range [see Fig. 5(c)]. The beam quality parameter was measured to be $M^2 = 1.2 \times 1.2$, assuming the central wavelength of 878 nm, thus showing the absence of beam degradation when propagating through cascaded multipass stages.

Additionally, we characterized spatio-spectral homogeneity by measuring spectra in different positions of the compressed beam, similar to stage 1 output. The overlap parameter V was used to quantify the measurement [Fig. 5a]. A perfect spectral overlap of > 99 % was measured in the beam's central part (defined as $1/e^2$) while going down to 90 % at the edges. The weighted average values of overlap factors were $V_x = 99.4$ % and $V_y = 99.4$ %, respectively. The results indicated an excellent spectral content enclosure over the beam area.

The power stability measurement [Fig. 5(b)] was performed over two hours of continuous operation. The system ran stably without a drop in output power.

To investigate the self-compression regime and enhance the spectral broadening, we reduced the Argon pressure slightly below 7.8 bar. As a result, the output spectrum spanned the range from 700-1350 nm (at -30 dB level), corresponding to an FTL of 5.6 fs [see Fig. 4(c, d)]. The output pulses were compressed with a pair of dispersive mirrors (2 x -45 fs$^2$) and a 5 mm AR-coated fused silica plate to 6.2 fs, thus crossing the mark of two optical cycles. Based on the retrieval, the main peak included 65 % of the pulse energy. In this case, we attribute a lower quality of pulse compression to our dispersive mirrors, which clearly could not support the full bandwidth of the output spectrum. The phase jump around the central wavelength [see Fig. 4(c)] reproducibly appeared in the retrieval when we operated stage 2 in the negative dispersion regime. All the spectra in this work are summarized in Fig. 6.

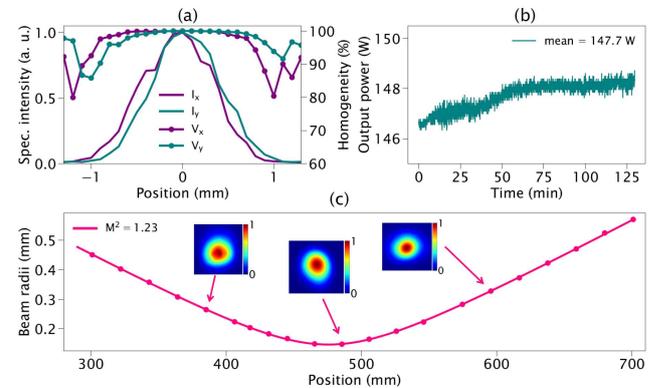

**Fig. 5.** Characterization of the stage 2 output with an FTL of 7.8 fs. (a) Spatio-spectral homogeneity $V_x = 99.4$ % and $V_y = 99.4$ %) and relative intensity I for tangential and sagittal planes. (b) Long-term measurement of the output power. (c) Measurement of the beam quality $M^2$. The central wavelength of 878 nm is an average weighted according to the sensitivity of the sensor.

We ran 1D simulations of pulse propagation through a quasi-waveguide filled with a nonlinear material in an open-source software PyNLO [28], similar to the previous work [14]. The simulations considered the optical properties of Argon gas and the cell mirror coatings. The experimental results agree with the numerical simulations [see Fig. 4(a, c)].

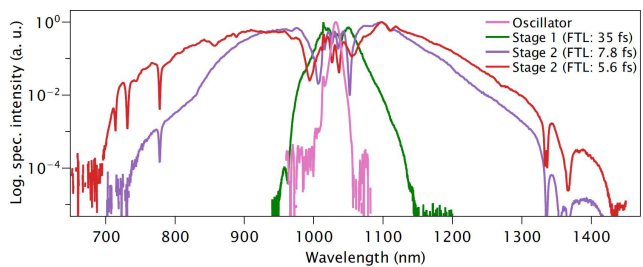

**Fig. 6**. Output spectra in the setup. The curves are measured with an optical spectrum analyzer. The pink curve is the oscillator spectrum, the green and purple curves correspond to the output of stage 1 and stage 2, respectively. The red curve was measured when stage 2 operated in an anomalous dispersion regime.

In conclusion, we demonstrated a compact dual-stage Herriott-type multipass system, which compressed 120 fs input pulses to 8.0 fs with 75 % energy in the main peak and 148 W average power corresponding to 0.9 GW peak power. In other words, when focusing the output down to a 30-40 µm spot in diameter, an intensity of ~$10^{14}$ W/cm$^2$ can be reached, sufficient for doing high harmonic generation in Argon or Krypton gases. Additionally, we reached a sub-two-cycle regime by operating the second stage in an anomalous dispersive regime, namely, 6.2 fs with a 65 % energy concentrated in the main peak and, thus, showed the feasibility of getting close to 5 fs with better dispersion management. The setup relying on the all-dielectrically coated mirrors and gas as a nonlinear medium proved highly suitable for spectral broadening and compression of high average- and peak-power Yb-based lasers. The laser system is an ideal high repetition rate driver for attosecond pulse generation, high harmonics generation-based sources, and XUV frequency comb spectroscopy.

**Acknowledgments.** We thank Christian Franke for designing the optomechanical components and cell housings. We also thank Moinuddin Kadiwala for cleaning the housings, providing optics, and inventorying them ☺.

**Disclosures.** OP and KF declare a conflict of interest as co-founders of n2-Photonics.

**Data availability.** Data underlying the results presented in this paper are available on request.

## References

1. G. Sansone, E. Benedetti, F. Calegari, C. Vozzi, L. Avaldi, R. Flammini, L. Poletto, P. Villoresi, C. Altucci, R. Velotta, S. Stagira, S. De Silvestri, and M. Nisoli, Science **314**, 443 (2006).
2. A. Pashkin, A. Sell, T. Kampfrath, and R. Huber, New J. Phys. **15**, 065003 (2013).
3. M. Plankl, P. E. Faria Junior, F. Mooshammer, T. Siday, M. Zizlsperger, F. Sandner, F. Schiegl, S. Maier, M. A. Huber, M. Gmitra, J. Fabian, J. L. Boland, T. L. Cocker, and R. Huber, Nat. Photon. **15**, 594 (2021).
4. P. B. Corkum and F. Krausz, Nat. Phys. **3**, 381 (2007).
5. R. Klas, A. Kirsche, M. Gebhardt, J. Buldt, H. Stark, S. Hädrich, J. Rothhardt, and J. Limpert, PhotoniX **2**, 4 (2021).
6. S. M. Teichmann, F. Silva, S. L. Cousin, M. Hemmer, and J. Biegert, Nat. Commun. **7**, 11493 (2016).
7. J. Schulte, T. Sartorius, J. Weitenberg, A. Vernaleken, and P. Russbueldt, Opt. Lett. **41**, 4511 (2016).
8. A.-L. Viotti, M. Seidel, E. Escoto, S. Rajhans, W. P. Leemans, I. Hartl, and C. M. Heyl, Optica **9**, 197 (2022).
9. M. Kaumanns, D. Kormin, T. Nubbemeyer, V. Pervak, and S. Karsch, Opt. Lett. **46**, 929 (2021).
10. M. Müller, J. Buldt, H. Stark, C. Grebing, and J. Limpert, Opt. Lett. **46**, 2678 (2021).
11. C. Grebing, M. Müller, J. Buldt, H. Stark, and J. Limpert, Opt. Lett. **45**, 6250 (2020).
12. A.-L. Viotti, C. Li, G. Arisholm, L. Winkelmann, I. Hartl, C. M. Heyl, and M. Seidel, Opt. Lett. **48**, 984 (2023).
13. L. Daniault, Z. Cheng, J. Kaur, J.-F. Hergott, F. Réau, O. Tcherbakoff, N. Daher, X. Délen, M. Hanna, and R. Lopez-Martens, Opt. Lett. **46**, 5264 (2021).
14. S. Goncharov, K. Fritsch, and O. Pronin, Opt. Lett. **48**, 147 (2023).
15. S. Rajhans, E. Escoto, N. Khodakovskiy, P. K. Velpula, B. Farace, U. Grosse-Wortmann, R. J. Shalloo, C. L. Arnold, K. Põder, J. Osterhoff, W. P. Leemans, I. Hartl, and C. M. Heyl, Opt. Lett. **48**, 4753 (2023).
16. P. Rueda, F. Videla, T. Witting, G. A. Torchia, and F. J. Furch, Opt. Express **29**, 27004 (2021).
17. V. Pervak, I. Ahmad, M. K. Trubetskov, A. V. Tikhonravov, and F. Krausz, Opt. Express **17**, 7943 (2009).
18. G. Steinmeyer, Opt. Express **11**, 2385 (2003).
19. A. Wirth, M. Th. Hassan, I. Grguraš, J. Gagnon, A. Moulet, T. T. Luu, S. Pabst, R. Santra, Z. A. Alahmed, A. M. Azzeer, V. S. Yakovlev, V. Pervak, F. Krausz, and E. Goulielmakis, Science **334**, 195 (2011).
20. A. Cingöz, D. C. Yost, T. K. Allison, A. Ruehl, M. E. Fermann, I. Hartl, and J. Ye, Nature **482**, 68 (2012).
21. D. Z. Kandula, C. Gohle, T. J. Pinkert, W. Ubachs, and K. S. E. Eikema, Phys. Rev. Lett. **105**, 063001 (2010).
22. J. Rothhardt, S. Hädrich, J. Delagnes, E. Cormier, and J. Limpert, Laser & Photonics Reviews **11**, 1700043 (2017).
23. S. Hädrich, M. Kienel, M. Müller, A. Klenke, J. Rothhardt, R. Klas, T. Gottschall, T. Eidam, A. Drozdy, P. Jójárt, Z. Várallyay, E. Cormier, K. Osvay, A. Tünnermann, and J. Limpert, Opt. Lett. **41**, 4332 (2016).
24. J. Drs, J. Fischer, N. Modsching, F. Labaye, M. Müller, V. J. Wittwer, and T. Südmeyer, Laser & Photonics Reviews **17**, 2200258 (2023).
25. J. Weitenberg, A. Vernaleken, J. Schulte, A. Ozawa, T. Sartorius, V. Pervak, H.-D. Hoffmann, T. Udem, P. Russbüldt, and T. W. Hänsch, Opt. Express **25**, 20502 (2017).
26. S. Goncharov, K. Fritsch, and O. Pronin, Opt. Express **31**, 25970 (2023).
27. E. R. Peck and D. J. Fisher, J. Opt. Soc. Am. **54**, 1362 (1964).
28. G. Ycas, "PyNLO's documentation," https://pynlo.readthedocs.io.